\documentclass[onecollarge]{svjour2}
\usepackage[sort&compress]{natbib}
\bibpunct{[}{]}{,}{n}{}{,}
\smartqed 
\usepackage{epsfig,graphicx,color}

\journalname{Few-Body Systems}

\begin{document}
\title{
Universality in the neutron$-^{19}$C scattering using finite-range separable interactions
 \thanks{Presented at the 23rd European Conference on Few-Body Problems in Physics, Aarhus, Denmark, August 2016}
}

\titlerunning{
Universality in the neutron$-^{19}$C scattering using finite range separable interactions}
\author{M. A. Shalchi         \and
        M.T. Yamashita \and
        M.R. Hadizadeh \and
        T. Frederico \and
        Lauro Tomio}

\authorrunning{Hadizadeh et al.}

\institute{ 
M.A. Shalchi,  M.T. Yamashita and Lauro Tomio \at
Universidade Estadual Paulista (Unesp), Instituto de Física Teórica (IFT),  
01140-070, S\~ao Paulo, SP, Brazil \\
\email{shalchi@ift.unesp.br}
\and
M. R. Hadizadeh \at
Institute for Nuclear and Particle Physics and Department of Physics and Astronomy, Ohio University, 
Athens, OH 45701, and
College of Science and Engineering, Central State University, Wilberforce, OH 45384, USA 
\and             
T. Frederico and Lauro Tomio 
\at Instituto Tecnol\'ogico de Aeron\'autica, DCTA, 12228-900, S\~ao Jos\'e dos Campos, Brazil  \\
}
\date{Received: date / Accepted: date}
\maketitle
\begin{abstract}
{
We report a study on the low-energy properties of the elastic $s-$wave scattering of a neutron ($n$) in the carbon
isotope $^{19}$C near the critical condition for the occurrence of an excited Efimov state in the three-body 
$n-n-^{18}$C system. For the separation energy of the two halo neutrons in  $^{20}$C we use the 
available experimental data. We also investigate to which extent the universal scaling laws, strictly valid in the zero-range limit, 
will survive when using finite-range interactions. By allowing to vary the $n-^{18}$C binding energy, a
scaling behavior for the real and imaginary parts of the $s-$wave phase-shift $\delta_0$ is verified,
emerging some universal characteristics given by the pole-position of $k\cot(\delta_0^R)$ and 
effective-range parameters.}
\end{abstract} 
\keywords{Halo nuclei, scattering theory, Efimov physics, Faddeev equation, Few body}
\vspace{.5cm}
{
\noindent {\bf Introduction - }
In the present communication we are reporting some results obtained when studying the elastic $s-$wave
scattering of a neutron from $^{19}$C, in the low-energy region, by considering a three-body $n-n-$core
system, where the core is given by $^{18}$C, with the binding energy of $^{20}$C fixed to 3.5 MeV.
Our aim is to investigate possible universal characteristics~\cite{FrePPNP12} obeyed by scattering observables 
of a three-body system which is close to form an excited Efimov state~\cite{Efimov}.
With this purpose, we notice that the structure of exotic halo nuclei has been shown quite fascinating 
to study when looking for universal characteristics in mass imbalanced systems~\cite{FrePPNP12,riisager2013}.
This motivated us to extend the corresponding studies from negative energy to the scattering
low energy sector, in order to evidence universal effects in the continuum which could be associated 
to the Efimov physics. Therefore, we are moving from the bound $n-n-c$ system to the case that a single 
neutron is scattered by the bound $n-c$ subsystem. 
As an appropriate interesting example, with enough experimental data to be considered in
theoretical models, we can mention the $^{20}$C nucleus within the three-body $n-n-^{18}$C 
model~\cite{YamNPA04}. This neutron-rich exotic nucleus, with the $n-^{18}$C subsystem forming 
an $s-$wave bound state, is close to produce an Efimov excited state~\cite{FedorovPRL94}.
Next, to verify possible universal properties which can be attributed to scattering observables,
we allow the two-body $n-^{18}$C bound-state energy to vary.

Within our present approach, we found convenient to remind some original studies considering 
the three-nucleon elastic scattering, given by neutron-deuteron ($n-d$) system. In the study of 
the spin-doublet $s-$wave $n-d$ elastic scattering, a relevant quantity that was considered is 
$k\cot\delta_0$, where $k$ is the on-energy shell relative momentum (incoming and final) and 
$\delta_0$ is the $s$-wave phase shift. It was first pointed out in Ref.~\cite{vanoers} that this 
quantity presents a pole when analyzing experimental data for the $n-d$ doublet $s-$wave phase 
shift, which have also motivated some other pioneering related works~\cite{reiner,wfuda,gfuda}.

The zero-range results reported in \cite{YamPLB08b} for the $n-^{19}$C elastic $s-$wave 
phase-shifts support the existence of a pole in the effective range expansion in a good 
qualitative agreement with what is found for the $n-d$ low-energy scattering, as well as for 
the atom-dimer discussed above. It remains the question how much such universal features 
are preserved when using finite-range potentials, considering the range of the $n-c$ and 
$n-n$ interactions, when the two-neutron separation energy in $^{20}$C is kept fixed.  
Therefore, in a recent work \cite{recent-paper},
we have performed a deeper investigation on the behavior of the low-energy pole 
$ k\cot\delta_0$ for the $n-^{19}$C scattering, beyond the zero-range approach 
used in Ref.~\cite{c20}, by considering a finite-range potential, which was chosen separable 
with Yamaguchi form.
The effect of mass asymmetry associated with finite-range interactions was also verified in
this study.

By reporting the main results given in Ref.~\cite{recent-paper}, where the formalism is detailed,
we found appropriate in this communication to introduce an extended analysis of our results on
the $n-^{19}$C elastic cross sections computations. The analysis is done with the assumption
that the required separation energy for two neutrons in $^{20}$C is fixed to an accepted 
experimental value, $E_{nnc}\equiv E_{^{20}{\rm C}}=-3.5$ MeV. Then, the less known two-body 
bound-state energy $E_{nc}\equiv E_{^{19}{\rm C}}$ is varied within our numerical investigation, 
looking for some characteristic universal behavior of an observable related to the $s-$wave scattering 
phase-shift $\delta_0$. Motivated by the well-known studies related to $n-d$, for such
observable we choose the energy position of the poles verified in $k\cot{\delta_0^R}$ (where
 ${\delta_0^R}$ is the real part of ${\delta_0}$). 
As this observable is expected to be quite sensible to the range, we found also appropriate to 
further discuss here some relevant results related to range effects and corresponding scaling
reported in Ref.~\cite{recent-paper}.  We have noticed that the scaling of the position of the pole 
in $k\cot{\delta_0^R}$ and the effective range parameters reproduces universal characteristics 
already found within the zero-range model under the realm of the Efimov physics.
 \vspace{0.5cm}

\noindent {\bf Formalism and notation - }
For the formalism, we follow the standard one for the elastic scattering amplitude of a neutron in the bound 
neutron-core subsystem, which is detailed in Refs.~\cite{c20,recent-paper}.
For the two-body $n-n$ and $n-c$ interactions, we consider separable expressions, 
with strengths and ranges adjusted by the $n-n$ virtual-state energy ($E_{nn}= -$143 keV) and 
$n-c$ bound-state energy ($E_{nc}=E_{^{19}{\rm C}}$), respectively, such that
 the three-body ground-state binding energy remains fixed to ($E_{^{20}{\rm C}}=-$ 3.5 MeV).
The neutron separation energy in $^{19}$C has a sizable error, with given values ranging from 
$-$160$\pm$110 keV~\cite{audi} to $-$530$\pm$130 keV~\cite{naka99}. 
Therefore, in our present study, a wide variation for $E_{nc}$ will be allowed, from 
$-$400 up to $-$800 keV, such that all possible experimental values can be included. 
For the $n-n$ and $n-c$ interactions, considering range-parameters given by $\beta_{nn}$ and
$\beta_{nc}$, respectively, we assume they have a one-term separable form, given by
\begin{eqnarray}\label{pot}
 V_{ij}(p,p')=\lambda_{ij}\left(\frac{1}{p^2+\beta_{ij}^2}\right)\left(\frac{1}{p'^2+\beta_{ij}^2}\right)
 \;\;\;  (ij=nn,\;nc).
\end{eqnarray} 
For negative two-body energies (bound or virtual), $E_{ij}$, defining $\kappa_{ij}\equiv \sqrt{2\mu_{ij} |E_{ij}|}$
with $\mu_{ij}$ being the reduced masses for $n-n$ and $n-c$ subsystems, 
$\mu_{nn}\equiv m_n/2$ and $\mu_{nc}\equiv m_n A/(A+1)$, where $A\equiv m_c/m_n$, 
the relations for the  strengths and ranges are, respectively,  given by
\begin{eqnarray}
 \lambda^{-1}_{ij}=\frac{-2\pi\mu_{ij}}{\beta_{ij}(\beta_{ij}\pm\kappa_{ij})^2},\;\;\;
 r_{ij}=\frac{1}{\beta_{ij}}+\frac{2\beta_{ij}}{(\beta_{ij}\pm\kappa_{ij})^2},\;\;\; (+/-)\,{\rm for\; bound/virtual},
\end{eqnarray}
Our units are such that $\hbar=1$, with momentum variables in fm$^{-1}$ and the unit conversion given by $\hbar^2/m_n= 
41.47$ MeV fm$^2$.

The on-energy-shell incoming and final relative momentum are related to the three-body energy 
$E \equiv E(k_i)$ by $k\equiv k_i\equiv |\vec k_i|=|\vec k_f| =  \sqrt{2\mu_{n(nc)}\left(E-E_{nc}\right)}$, 
where $\mu_{n(nc)}=m_n(A+1)/(A+2)$. 
The coupled $s-$wave scattering equations can be cast in a single channel Lippmann-Schwinger-type 
equation for the relevant amplitude $h_n$, with the following form (for details, see 
Ref.~\cite{recent-paper}):
\begin{eqnarray}
h_n(q,E)&=&  \mathcal{V} \,(q,k,E)+\frac{2}{\pi}\int_0^\infty dq' \, q'^2  \, \frac{\mathcal{V} 
\,(q,q',E)h_n(q',E)}{q'^2-k^2-{\rm i} \epsilon},\end{eqnarray}
where the kernel $\mathcal{V}(q,q',E)$ for the  $n-(nc)$ channel amplitude contains the contribution of the 
coupled $(nn)-c$ channel. 

\noindent {\bf Results and discussion -}
The neutron-core interaction-range parameter, $\beta_{nc}$, of the one-term separable potential, is varied for different $E_{^{19}{\rm C}}$, by keeping fixed the low-energy observables $S_{2n}$ and the virtual state energy $E_{nn}$. For that, we study situations with two distinct regions, with low and high values for  the range parameters $\beta_{nn}$ and $\beta_{nc}$, which  will correspond to
 larger and smaller effective ranges of the $n-n$ and $n-c$ interactions, respectively. 
 Here we include some representative results, which are relevant for our following discussed on the scaling characterization 
 of the low-energy neutron scattering in  $^{19}C$. The results are resumed in Table \ref{table1} and Fig.~\ref{fig1}. 
\begin{table}[h]
  \caption{One-neutron separation energy in $^{19}$C (left column), obtained by the given range parameters of the separable Yamaguchi potential, $\beta_{nc}$ (second and fourth columns), with the corresponding effective ranges (third and fifth columns).
The values of $\beta_{nc}$ were obtained by fitting the two-neutron separation energy in $^{20}$C (3.5 MeV~\cite{audi}), with the 
$n-n$ interactions fixed by the virtual state energy, $E_{nn}= -$143 keV.
 The fixed $\beta_{nn}$ values for high and low ranges, respectively, are 1.34 fm$^{-1}$ ($r_{nn}=2.372$ fm) and 
 24.50 fm$^{-1}$ ($r_{nn}=$0.1228 fm).}
 \centering
 \label{table1}
\begin{tabular}{ccccc}
\hline
$|E_{^{19}{\rm C}}| $(keV)& $\beta_{nc} ({\rm fm}^{-1})$  & $r_{nc}$(fm)& $\beta_{nc}({\rm fm}^{-1})$&$r_{nc}$(fm)   \\
\tableheadseprule\noalign{\smallskip}
 200& 0.971 & 2.736 &18.970& 0.157\\
 400& 0.754 & 3.233 & 17.036& 0.174\\
 600&0.598  & 3.720 & 15.592& 0.190\\
 800& 0.477 & 4.255 &14.395& 0.205\\
\hline\hline
\end{tabular}
 \end{table}

  \begin{figure}[h]
\begin{center}
\includegraphics[scale=0.6]{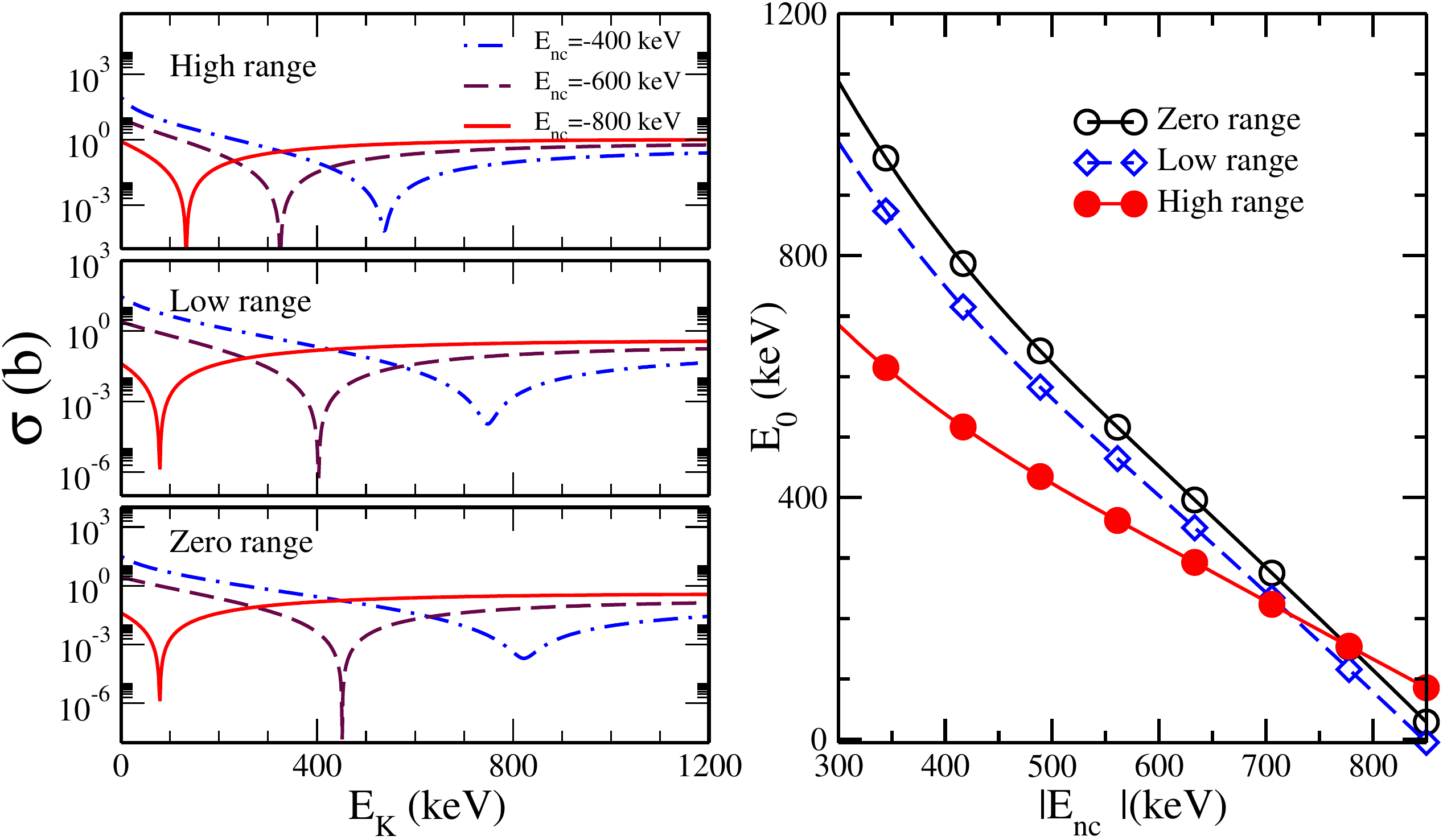}
\end{center}
\caption{
\textbf{Left panels:}
Several plots for the $s-$wave $n-\,^{19}$C elastic  cross-section are shown as functions of the center-of-mass kinetic energy $E_K$.
In each panel we have three plots corresponding to different $E_{nc}$ binding energies, which are identified in the top panel.
For the range parameters used in the Yamaguchi potential, with $E_{^{20}{\rm C}}$ fixed, high range are in the top panel;
low-range in the middle panel; and zero-range in the bottom panel.
\textbf{Right panel:} The energy positions of the $kcot\delta_0$ poles, given by $E_0$, are shown as the $n-^{18}$C binding energy
($E_{nc}$) is changed, considering results obtained with zero-, low- and high-range potential parameters.}
\label{fig1}
\end{figure}

In Table~\ref{table1}, by keeping fixed the three-body bound-state energy, $E_{^{20}{\rm C}}=-$3.5 MeV, as well as the $n-n$ 
parameters, such that $E_{nn}=$-143 keV, we report a few values for the one-neutron separation energies in $^{19}$C, given by $E_{^{19}{\rm C}}$, with the respective values for the parameter $\beta_{nc}$ and corresponding effective ranges, given by $r_{nc}$. For the $n-n$ parameters, we have fixed $\beta_{nn}=1.34$ fm$^{-1}$ for the high-range cases, and $\beta_{nn}=24.50$ fm$^{-1}$ for the low-ranges ones.
For lower values of $\beta$'s, the effective range of the $n-c$ interaction is comparable to the $^{18}$C root-mean-square 
matter radius, $r_m[^{18}C]$= 2.82$\pm$0.04 fm \cite{OzaNPA01}, while for high values of $\beta$'s, $r_m[^{18}C]$ is quite 
large compared to the effective-range, being not realistic. The case when both $\beta_{nn}$ and $\beta_{nc}$ are large, the corresponding 
effective ranges become much smaller than the core size and the $n-n$ effective range, $2.75$ fm \cite{MilPRP90}.

The energy positions $E_0$ of the $k\cot\delta_0^R$ poles are plotted in the right-side panel of Fig.~\ref{fig1} 
as a function of the corresponding $n-c$ binding energies, for the zero-range, high-beta (low-range) and low-beta 
(high-range) parameters. As shown, $E_0$ obtained with high-range parameters are less affected by the changes 
of $E_{nc}$ than the other two cases. The quite different slope of the curves obtained with high-range parameters
when compared with the corresponding ones with low- or zero-range parameters can also be observed in the
results presented in the left-hand-side panels of Fig.~\ref{fig1}.
 
In the panels at the left side of Fig.~\ref{fig1} we have the results for the $s-$wave scattering cross sections (in
logarighmic scale) obtained from $d\sigma/d\Omega$ $= \left| h_n(k;E)\right|^2$, given as functions of the 
center-of-mass kinetic energy $E_K$. There are three panels corresponding, respectively, to high-, low- and 
zero-range parameters. In each panel we show three representative values for the $n-c$ binding energy 
($E_{^{19}C}=-$400, $-$600 and  $-$800 keV). The poles of $k\cot\delta_0^R$ correspond to the observed 
minima  appearing  in the cross-section for  $E_K=E_0$, clearly seen for $E_0$ below the breakup threshold or 
when the absorption to the breakup channel is not large. 
In this trend of moving the poles with the variation of $E_{^{19}{\rm C}}$ we should notice that such behavior 
affects more the low-range cases, which reflects the slope changes verified in the right side of Fig.~\ref{fig1}.

We finally should observe that the poles are correlating the $s-$wave quantities with the binding energy of 
the three-body halo system, namely the $^{20}$C nucleus. The strong correlation, previously verified using
zero-range interactions, survives when considering finite-range potential with effective $n-n$ and 
$n-c$ ranges compatible with physical values.  

\begin{acknowledgements}
This work was partly supported by funds provided by the Brazilian agencies Coordena\c c\~ao de Aperfei\c coamento de Pessoal
de N\'\i  vel Superior - CAPES [Proc. no. 88881.030363/2013-01(MTY and MAS) and a Senior Visitor Program at the Instituto 
Tecnol\'ogico de Aeron\'autica (LT)], Conselho Nacional de Desenvolvimento Cient\'\i  fico e Tecnol\'ogico - CNPq [grants no. 302701/2013-3(MTY), 306191/2014-8(LT), 308486/2015-3 (TF)]
and Funda\c c\~ao de Amparo \'a Pesquisa do Estado de S\~ao Paulo - FAPESP [grant no. 2016/01816-2(MTY)].
 M.R.H. acknowledges the partial support by National Science Foundation under Contract No. NSF-HRD-1436702 with 
 Central State 
 University and by the Institute of Nuclear and Particle Physics at Ohio University.
 \end{acknowledgements}
  }


\vspace{-0.7cm}

\begin{thebibliography}{3}
\bibitem{FrePPNP12} 
T. Frederico, A. Delfino, L. Tomio, M.T. Yamashita, Prog. Part. Nucl. Phys. {\bf 67} (2012) 939.

\bibitem{Efimov} V. Efimov, Phys. Lett. B {\bf 33} (1970) 563.

\bibitem{riisager2013} K. Riisager,  Phys. Scr. {\bf T152} (2013) 014001.

\bibitem{YamNPA04}  M.T. Yamashita, L. Tomio and T. Frederico, Nucl. Phys. A {\bf 735} (2004) 40.

\bibitem{FedorovPRL94} D.V. Fedorov, A.S. Jensen and K. Riisager, Phys. Rev. Lett. {\bf 73} (1994) 2817.
\bibitem{vanoers} W.T.H. van Oers and J.D. Seagrave, Phys. Lett. B {\bf 24} (1967) 562.
 \bibitem{reiner} A.S. Reiner, Phys. Lett. B {\bf 28} (1969) 387.
\bibitem{wfuda} J.S. Whiting and M.G. Fuda, Phys. Rev. C {\bf 14} (1976) 18.
\bibitem{gfuda} B.A. Girard and M.G. Fuda, Phys. Rev. C {\bf 19} (1979) 579.
\bibitem{YamPLB08b}  M.T. Yamashita, T. Frederico and L. Tomio, Phys. Lett. B {\bf 670} (2008) 49.
\bibitem{recent-paper} M.A Shalchi, M. T. Yamashita, M. R. Hadizadeh, T. Frederico, L. Tomio, 
Phys. Lett. B, {\bf 764} (2017) 196.
\bibitem{c20}  M.T. Yamashita, T. Frederico and L. Tomio, Phys. Lett. B {\bf 660} (2008) 339.
\bibitem{audi}
G. Audi, A.H. Wapstra, and C. Thibault, Nucl. Phys. A {\bf 729} (2003) 337.
\bibitem{naka99} T. Nakamura et al., Phys. Rev. Lett. {\bf 83} (1999) 1112.
\bibitem{OzaNPA01} A. Ozawa et al., Nucl. Phys. A {\bf 691}, 599 (2001).
\bibitem{MilPRP90} G. A. Miller, M. K. Nefkens, and I. Slaus, Phys. Rep. {\bf 194} (1990) 1.
\end{thebibliography}
\end{document}